\documentclass[%
 aip,
 jmp,%
 amsmath,amssymb,
preprint,%
]{revtex4-1}

\usepackage{graphicx}
\usepackage{dcolumn}
\usepackage{bm}
\usepackage{color}
\usepackage{epstopdf}

\begin{document}

\preprint{AIP/123-QED}

\title[Rheology of sediment transported by a laminar flow]{Rheology of sediment transported by a laminar flow}
\author{M. Houssais}
 \email{housais.morgane@gmail.com.}
 \affiliation{Department of Earth and Environmental Science, University of Pennsylvania.
}%
 \author{C. P. Ortiz}%
  \affiliation{Department of Earth and Environmental Science, University of Pennsylvania.}
\affiliation{ Department of Physics and Astronomy, University of Pennsylvania,
}%
\author{D. J. Durian}
\affiliation{ Department of Physics and Astronomy, University of Pennsylvania,
}%
\author{D. J. Jerolmack}
\affiliation{Department of Earth and Environmental Science, University of Pennsylvania.
}%

\date{\today}

\begin{abstract}

Understanding the dynamics of fluid-driven sediment transport remains challenging, as it is an intermediate region between a granular material and a fluid flow. Boyer \textit{et al.}\citep{Boyer2011} proposed a local rheology unifying dense dry-granular and viscous-suspension flows, but it has been validated only for neutrally-buoyant particles in a confined system. Here we generalize the Boyer \textit{et al.}\citep{Boyer2011} model to account for the weight of a particle by addition of a pressure $P_0$, and test the ability of this model to describe sediment transport in an idealized laboratory river. We subject a bed of settling plastic particles to a laminar-shear flow from above, and use Refractive-Index-Matching to track particles' motion and determine local rheology --- from the fluid-granular interface to deep in the granular bed. Data from all experiments collapse onto a single curve of friction $\mu$ as a function of the viscous number $I_v$ over the range $10^{-5} \leq I_v \leq 1$, validating the local rheology model. For $Iv < 10^{-5}$, however, data do not collapse. Instead of undergoing a jamming transition with $\mu \rightarrow \mu_s$ as expected, particles transition to a creeping regime where we observe a continuous decay of the friction coefficient $\mu \leq \mu_s$ as $I_v$ decreases. The rheology of this creep regime cannot be described by the local model, and more work is needed to determine whether a non-local rheology model can be modified to account for our findings.

\end{abstract}

\keywords{Rheology, soft condensed matter, sediment transport}
\maketitle


\section{\label{sec:level1}Introduction}



Sediment transport in rivers involves the entrainment and movement of a granular material by a unidirectional shear flow. Historically, research has emphasized the fluid dynamics: a fluid flow field over a rough static bed develops a characteristic shear stress $\tau$, which triggers the entrainment of grains at the bed surface above a critical value $\tau_c$. Numerous experimental, analytical and field studies have shown that the value of $\tau_c$ depends on the particle Reynolds number and the bed-surface granulometry \citep{Shields1936, Wiberg1987, Buffington1997, Wilcock2003, Houssais2012}. This work has been integral to the development of equations for predicting rates of sediment transport. Much research has focused on bed-load transport \citep{Meyer-Peter1948, Lajeunesse2010} --- the movement of grains by rolling, sliding and hopping along the river bed --- because of its importance for shaping ripples and dunes \citep{Charru2004, Devauchelle2010}, and for determining river channel geometry \citep{Parker1978, Paola1992, Seizilles2013}. However, there are three major limitations to this framework. First, the processes of bed load, suspension, landslides and hillslope-soil creep are considered distinct and they are described by different transport relations. Yet, all of these processes involve the movement of grains by a tangential stress, and sediment transport transitions continuously across these processes in the landscape. Second, the threshold of sediment transport has been observed to vary through time\citep{Turowski2011, Hsu2011}, violating the classical prediction of a unique $\tau_c$ value for a given system; similar behavior has been seen for the case of landslide triggering due to heavy rain\citep{Iverson2000}. Third, empirical sediment-transport laws notoriously break down as the shear stress enters the vicinity of the critical value $\tau_c$ \citep{Recking2012}.

Although turbulence is often cited as the culprit for the apparent complexity of river sediment transport\citep{Diplas2008, Schmeeckle2014}, experiments and theory have shown that bed load in laminar flows is similar in many respects to its turbulent counterpart \citep{Charru2004, Malverti2008, Houssais2012, Houssais2015}. Moreover, recent studies have emphasized the importance of granular interactions to accurately model sediment transport by fluid flows \citep{Frey2011, Ouriemi2009, Aussillous2013, Chiodi2014, RevilBaudard2015}. Finally, our previous work suggested that the bed-load regime of sediment transport behaves as a dense granular flow\citep{Houssais2015}. These studies provide evidence that explicit consideration of granular rheology may address the limitations of the current sediment transport framework. In this paper, we develop the granular rheology approach to sediment transport, and explicitly test its ability to describe various regimes of fluid-driven granular motion.

Considering rheology, the shear stress is $\tau = \eta_{eff} \dot{\gamma}$ where $\eta_{eff}$ is the effective viscosity and $\dot{\gamma}$ is the local shear rate. Granular flows exhibit a nonlinear rheology: for dry systems, it has been established experimentally that the effective viscosity decreases as the local shear rate increases relative to the local confinement pressure $P_p$ \citep{GDRMidi2004}. 
As dissipation in dense granular flows results from friction at the contact surfaces between particles, one would rather examine the local rheology in terms of the evolution of the friction coefficient: 
\begin{equation}
\mu = \frac{\tau}{P_p} \; .
\end{equation} 
All values of $\mu$ observed in sheared granular experiments collapse to a single function of a time-scale ratio $ \frac{t_{micro}}{t_{macro}}$ (refs. \citenum{GDRMidi2004, Jop2005, Jop2006, Forterre2008}). Here $t_{macro} \equiv 1/\dot{\gamma}$ is the average macroscopic timescale of system deformation, and $t_{micro}$ is the microscope timescale for particle rearrangement in the pack due to the confinement pressure. The timescale definition varies with the surrounding fluid properties, and was shown to depend principally on two other parameters, the ratio of particle and fluid densities $\rho_p/\rho$, and the Stokes number \citep{Courrech2003, Cassar2005}.

For the case of a granular material submerged in a fluid of viscosity $\eta_f$, in the limit $\rho_p/\rho \simeq 1$ and the Stokes number $St<1$, $t_{micro}$ is driven by viscous drag, therefore the dimensional analysis leads to: $t_{micro} = \eta_f/P_p$. The corresponding time scale ratio was defined as the Viscous number \citep{Courrech2003, Cassar2005}:
\begin{equation}
I_v = \frac{\eta_f |\dot{\gamma}|}{P_p} \;.
\end{equation}



In the submerged case, the shear stress $\tau$ is driven by fluid and particle effects, $\tau = \tau_p + \tau_f$. Accordingly, the effective friction coefficient $\mu$ results from the sum of the particle-particle and particle-fluid interactions. \citet{Boyer2011}  conducted shear-cell experiments using neutrally-buoyant particles ($\rho_p/\rho = 1$) immersed in a viscous fluid, subject to an imposed confining pressure. They proposed that the rheology of such a system can be described as a smooth transition between the dry-granular rheology and the fully-suspended rheology (where particle-particle contacts are completely neglected). Indeed, their measurements of bulk parameters remarkably showed:
\begin{equation}
\mu(I_v) =  \mu_{dry}(I_v)+ \mu_{susp}(I_v) = \mu_s + (\mu_d-\mu_s)/(I_0/I_v+1) + I_v + \frac{5}{2} \Phi_c I_v^{1/2} \; ,
\label{eq:muIv}
\end{equation}
where $I_0 = 0.005$, $\mu_s = 0.32$, and $\mu_d = 0.7$ are the classical constants observed for dry granular flow experiments \citep{GDRMidi2004, Jop2006}. The term $\Phi_c$ is the packing fraction at which viscosity diverges in suspension experiments, found equal to 0.585 by Boyer \textit{et al.}\citep{Boyer2011}.
The associated relationship of the bulk packing fraction $\Phi$ with the Viscous number was found to be:
\begin{equation}
\Phi = \frac{\Phi_c}{1+I_v^{1/2}} \; .
\label{eq:CIv}
\end{equation}

This suggests that an important range of sediment transport processes in nature --- landslides, bed load and suspension ---  could be modeled as a unique, highly non-linear, material flow. This formalism has been validated by Boyer \textit{et al.}\cite{Boyer2011} for uniformly confined and shear system.  However, it has not yet been tested for conditions found in natural riverbeds made of settling particles ($\rho_p/\rho > 1$) where the pressure and shear rate vary vertically.  Besides the possibility of non-local effects, other issues might be expected to arise near the sediment surface where the grain concentration and pressure both vanish and also deep into the bed where creep occurs\citep{Komatsu2001}.



\begin{figure}
\centerline{\includegraphics[width=450pt]{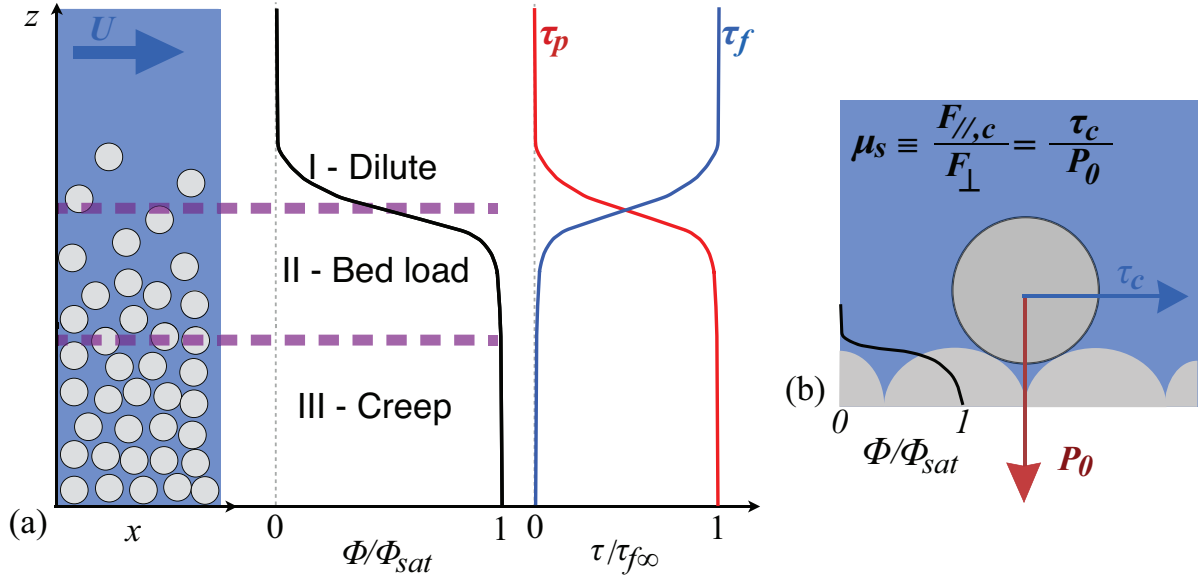}}
\caption{ Shematic representation of sediment transport. a) 2D sketch of sediment transport above the critical shear stress, $\tau>\tau_c$. Black, red and blue curves, respectively, show vertical trends of the packing fraction $\Phi$, the shear stress due to particles $\tau_p$, and the shear stress associated with the fluid $\tau_f$. $\tau_{f\infty}$ is the constant fluid-shear stress far from the particles. b) Schematic of entrainment of a single grain at the threshold of motion, on an idealized bed surface. In that case, by definition the friction coefficient is the static value $\mu_s$, strictly equal to the ratio of the tangential force $F_{//}$ to the normal force $F_{\perp}$ applied to the particle.}
\label{fig:SedTransportSketch}
\end{figure}

Our recent experimental results showed that settling particles entrained by a laminar fluid flow exhibit three different regimes of motion  as a function of depth into the bed\citep{Houssais2015}: (I) a dilute regime where particle motion is mostly driven by fluid-flow stress; (II) a denser particle flow, similar to a dry-granular flow, that we identified as bed-load; and (III) a creep regime associated with exceedingly slow and intermittent particle motion (see figure \ref{fig:SedTransportSketch}a). Particle velocity and concentration changed continuously across these regimes; however, the transition to creep occured at a fixed Viscous number, regardless of the applied fluid stress.  
Drawing on these observations, in this paper we determine the rheology of laminar sediment transport across all three regimes, and confront these results with the local rheology proposed by Boyer \textit{et al.}\citep{Boyer2011}.

\section{Experimental setup and methods}
\label{sec:setup}

\subsection{Experiment setup}

\begin{figure}
\centerline{\includegraphics[width=450pt]{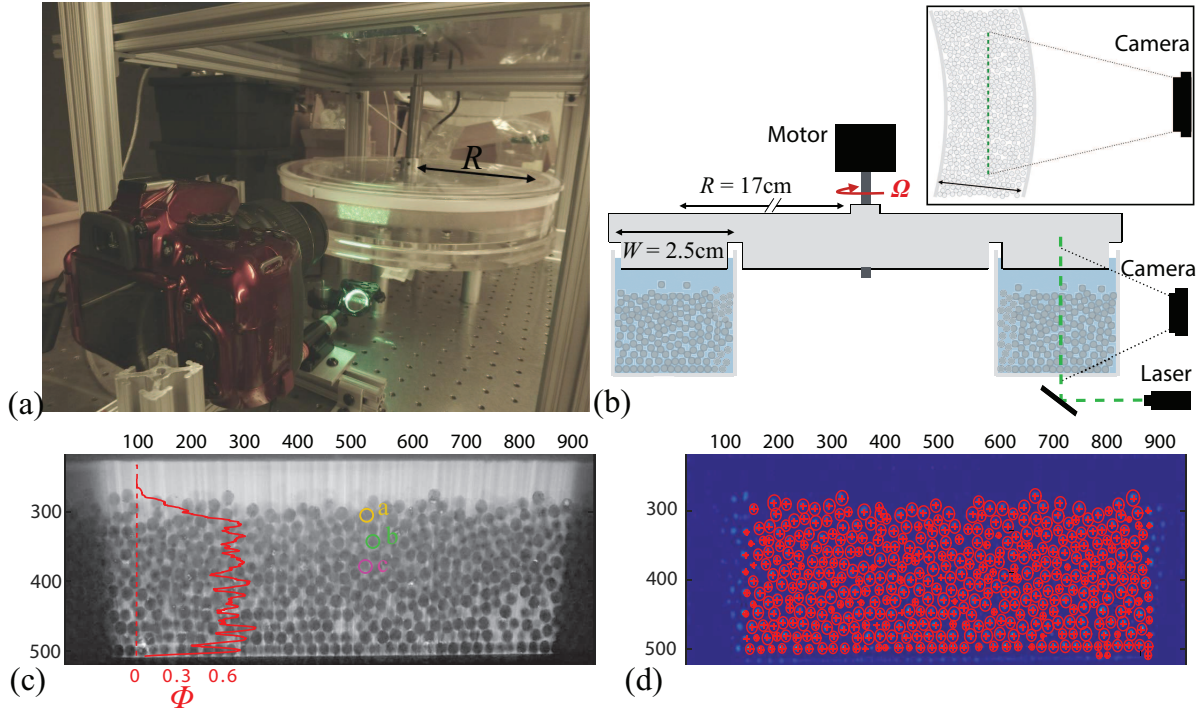}}
\caption{Experimental setup and particle detection. a) Picture of the experiment with camera, laser and an illuminated 2D plane of particles. b) Sketch of the experimental setup, with dimensions indicated. Insert: Top view.  c) Image showing 2D plane of particles. Red curve is the depth-varying packing fraction, computed at each elevation as the fraction of the image in the x-direction that is occupied by particles. The trajectories of particles highlighted here are shown in Fig. \ref{fig:particleTrajectories}. d) Typical particle detection result. Experimental results correspond to a run performed at $U = 48~\mathrm{mm/s}$.}
\label{fig:setup}
\end{figure}

Technical details of the experiments performed were presented in \citet{Houssais2015}, so we only briefly review them here. The setup consists of a closed annular flume of radius $R = 17~\mathrm{cm}$, in which we submerge a layer of spherical acrylic particles of diameter $d = 1.5~\mathrm{mm}$ and density $\rho_p =  1.19~\mathrm{g/mL}$ in an oil of viscosity $\eta_f = 68.6~\eta_{water}$ and density $\rho =  1.05~\mathrm{g/mL}$ (see figure \ref{fig:setup}). The system has width $W = 17d$, depth $H = 14d$, and is sheared by rotating the top of flume at a constant rate $\Omega$ (from 0.8 to $4~\mathrm{rpm}$) which corresponds to a top plate velocity, $U = 2 \pi R \Omega$ (from 14 and $48~\mathrm{mm/s}$). Below the plate is a fluid gap with a flow depth $h_f$, which is measured and ranges from 3.8 to $5.6~\mathrm{mm}$.  The low Reynolds numbers ($\mathrm{Re} = \rho h_f U/\eta_f \leq 3$)  
and low aspect ratio $h_f/W$ act to suppress turbulence and secondary flows \citep{Charru2004}, and allow us to measure the slow dynamics of particles as in an infinite, straight channel. To visualize granular dynamics, we index-match the PMMA particles with a viscous oil ($85~\%$ of PM550 and $15~\%$ of PM556 from Dow Corning, as previously used\citep{Stohr2003}), and record fluorescence of a dye (Exciton, pyrromethene 597) dispersed in the fluid and excited with a green laser sheet ($517~\mathrm{nm}$, $50~\mathrm{mW}$) of thickness $\simeq d/10$ (ref. \citep{Dijksman2012a}) aligned with the middle of the channel width (see figure \ref{fig:setup}b). Therefore, we image a vertical plane that is farthest from the influence of the side walls.

The  granular bed is prepared for each experiment with the following protocol: for 5 minutes the flume top is rotated at $3~\mathrm{rpm}$, applying a shear stress strong enough to suspend all particles, except the 2 last layers at the bottom which crystallize. The rotation then slowly returns to zero, and the particles settle for 5 minutes, building a random packed layer of approximately 11$d$. Then, a constant rotation $\Omega$ of the top plate drives the system during the entire experiment.
The duration of the experiment is not fixed; each lasts long enough (10 hours to several days) that all particles present in the recorded frames exhibit detectable displacement during the run.
With a single camera two different records are acquired: 20-min long movies with a frame rate of $30~\mathrm{Hz}$, able to capture particle flights at the surface, and hours-long time-lapse at $0.067~\mathrm{Hz}$, able to capture slow creeping motion deep inside the bed.

\subsection{Analysis}


To compute particle positions and apparent size, each recorded image is processed in the following manner.
First, a convolution with a disk of a radius close to that of a particle filters most of the image noise. Second, a radial symmetry analysis is made at each pixel, to reveal particle center positions as the most symmetrical objects. Finally, for each of these positions the average distance to the particle boundary (obtained from a binary version of the raw image) is taken as the apparent radius of the particle (see result example figure \ref{fig:setup}d).  

\begin{figure}
\centerline{\includegraphics[width=300pt]{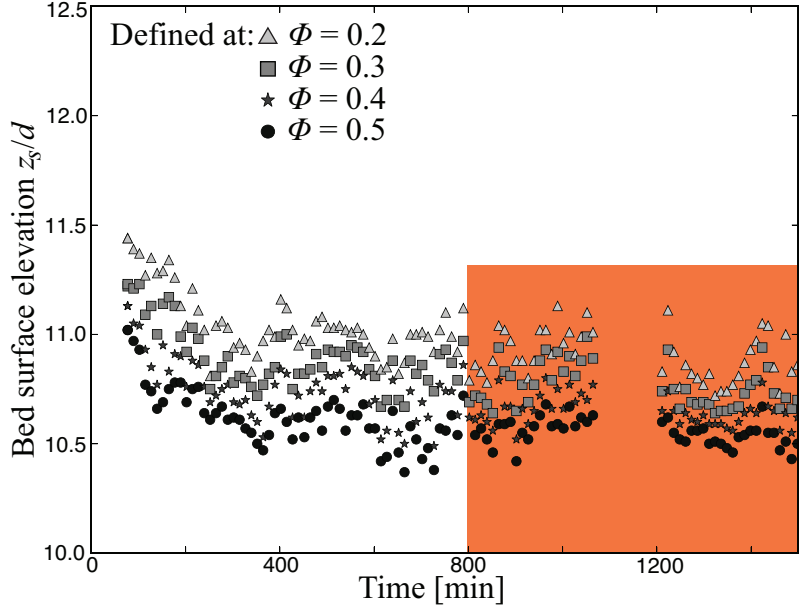}}
\caption{Time evolution of bed surface elevations defined at different specific packing fraction $\Phi$,  for  $U = 14~\mathrm{mm/s}$. The orange area represents the period over which long term measurement has been performed.}
\label{fig:ConcTimeEvol}
\end{figure}

Vertical profiles of particle concentration are computed from the image of the detected particle areas, by averaging pixels in the $x$ direction (see profile example figure \ref{fig:setup}c). 
We assume the particle concentration measured in a two-dimensional (2D) plane is a good proxy for the packing fraction $\Phi(z)$, as our measured saturation values deep inside the bed are close to classical values found for random packing fraction (0.58 to 0.6), and close to the value of $\Phi_c$ in equations (\ref{eq:muIv}) and (\ref{eq:CIv}).  
Each experiment exhibits an initial phase of fast compaction, which drives a temporal evolution of $\Phi$. Figure \ref{fig:ConcTimeEvol} presents the typical time evolution of the bed surface elevation. To study the steady-state rheology, we begin collecting data after most of the compaction has occurred (orange area on figure \ref{fig:ConcTimeEvol}).
 
Even after the compaction stage, we observe significant fluctuations from image to image due to averaging in a single vertical plane; however, the concentration profile always saturates to a constant value below a certain depth. We compute this saturation value (for $1<z/d<9$), $ \langle \Phi_{sat} \rangle _k$  for each experiment $k$ ($k = 1,5$), from time-averaged profiles made from hundreds to thousands of images. We find the mean value $\langle \Phi_{sat} \rangle_m = 0.589$, with slight variations (0.5 to $3~\%$ of $\langle \Phi_{sat} \rangle_m$) from experiment to experiment.  In order to ensure that packing fraction profiles for all experiments converge with each other at depth, we present  $k$-experiment profiles normalized by $\langle \Phi_{sat} \rangle_k/\langle \Phi_{sat} \rangle_m$.

To compute particle mean velocity, as in \citet{Houssais2015} we use Lagrangian particle tracking\citep{Ouellette2005}. 
From the particle tracks, we then compute individual velocities by measuring durations over which particles exhibit a displacement $\Delta x$ larger than our resolution limit $\delta x$ = $3~\mathrm{pixels}$. Final profiles of horizontal velocity are computed by averaging elevation strips in the $x$ direction over hours of records (see more details in Supplementary Information of \citet{Houssais2015}). 

The sediment bed is driven by a fluid and therefore there is no imposed confinement pressure; instead, there is a free-surface condition. As a consequence, the local pressure $P_p$ varies with depth due to the increasing overburden of particles. This can be computed from the packing fraction profile:
\begin{equation}
P_p (z)=(\rho_p - \rho) g \int_z^{+\infty} \langle \Phi \rangle (z) dz =  (\rho_p - \rho) g \Sigma(z) + P_0 \; ,
\label{eq:Pp}
\end{equation} 
with the gravity $g$, $\Sigma(z) = \sum \limits_z^\infty \langle \Phi \rangle(z) dz$ and $P_0$ is a constant of integration. Physically, the value of $P_0$ corresponds to the normal stress on a particle at elevations where $\langle \Phi \rangle$ becomes zero; it should relate to the weight of an individual particle. We don't attempt to account for the Janssen effect that might arise from the presence of side walls. The bed depth ($\simeq11d$) is smaller than the channel width ($17d$), however, so we expect that the confinement pressure should not saturate with depth \citep{Ovarlez2003}. The modified local pressure equation (\ref{eq:Pp}) we propose is novel in that it includes an explicit term for particle weight ($P_0$). Below we explore the consequences of the $P_0$ term for modeling the local rheology of sediment transport. 

\section{Results}
\label{sec:results}

\subsection{Phenomenology and shear stress measurement}

\begin{figure}
\centerline{\includegraphics[width=450pt]{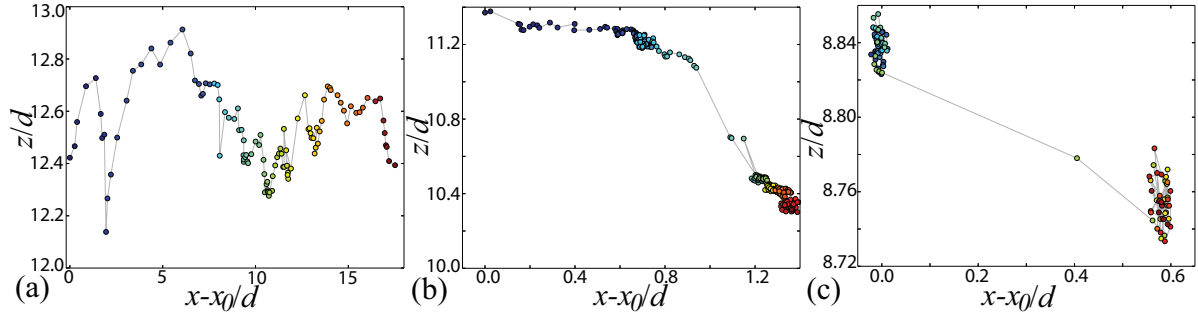}}
\caption{Example of trajectories $z(x)$ at different elevations in the sediment bed. a) $z \simeq  11d$, b) $z \simeq 9d$, c) $z \simeq  7d$ (see positions in figure \ref{fig:setup}c), for the experiment performed at $U = 48~\mathrm{mm/s}$, captured during the same 20min movie.  Color represents time, normalized by the total duration of each trajectory: a) 50~s, b) 115~s and c) 35~s, respectively. Note different axis scales for each plot. Consecutive points are all separated by 0.33~s.}
\label{fig:particleTrajectories}
\end{figure}

\begin{figure}
\centerline{\includegraphics[width=300pt]{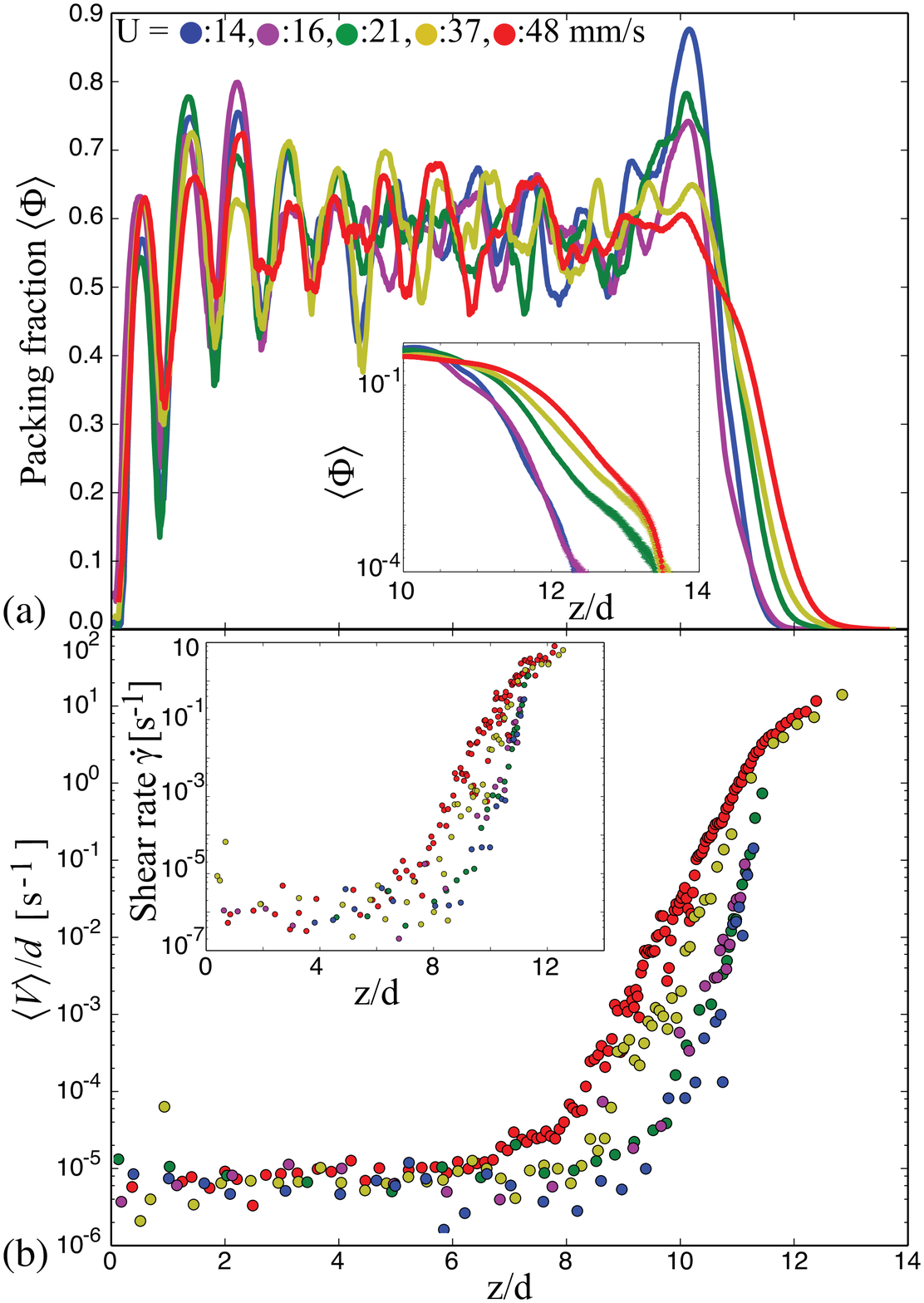}}
\caption{Depth-dependent concentration and velocity. a) Vertical profiles of long-time averaged particle concentration. Insert shows semilog plot, where error bars are visible. b) Vertical profiles of long-time averaged particle velocity. Insert presents its derivative, the particle shear rate $\dot{\gamma}$.}
\label{fig:CIvProfiles}
\end{figure}

For each experiment driven at a different rotation rate, we observe the same phenomenology: the particles at the bed surface are entrained by the fluid, and present classical features of rolling and saltation, with significant velocity oscillations \citep{Abbott1977, Charru2004, Lajeunesse2010} (see figure \ref{fig:particleTrajectories}a).  Particles just below the surface are also transported, due to grain and fluid motion above, but their trajectories remain confined as in a granular flow (see figure \ref{fig:particleTrajectories}b). Finally, particles deep inside the bed experience slow and sporadic motion that we identify as creep. Most the time these particles appear to be caged \citep{Reis2007}, but occasionally they make a fast but small displacement (see figure \ref{fig:particleTrajectories}c).
The range of stresses for our experiments were all low enough that entrained particles never touch the rotating top plate, which means that the concentration of particles always drop to zero at some height above the bed\citep{Houssais2015} (see figures \ref{fig:setup}c).
On figure \ref{fig:CIvProfiles}a and b are reported the long-time averaged profiles of packing fraction $\langle \Phi \rangle$ and streamwise velocity $\langle V \rangle$ obtained for 5 experiments performed at $U = 14, 16,  21,  37$ and  $48~\mathrm{mm/s}$. The concentration profiles all attain the saturation value $\langle \Phi_{sat} \rangle_m$ in the lowest part of the bed, and all always drop to zero moving up across the grain-fluid interface --- a distance of 2 to 3 particle diameters. For all stresses, one can observe that the velocity is smallest at the bottom, increases continuously with increasing $z$, and exhibits a significant kink at $\langle V \rangle/d \simeq 10^{-5}~\mathrm{s^{-1}}$; the depth associated with this kink varies with the flow speed. The two highest-flow velocity experiments present a second kink in the vicinity of the surface ($z/d \simeq 11$). 

The fluid depth $h_f$ is measured from $\langle \Phi \rangle$ profiles $h_f = H - z_s$, where $H$ is the total depth of the flume and $z_s$ is the elevation at which $\langle \Phi \rangle |_{z = z_s} = \langle \Phi_\mathrm{sat} \rangle_m/2$, an indicator of the bed surface\citep{Duran2012}. Therefore we compute the mean fluid shear stress in the region $z_s \leq z <H$ at steady state, and assume it to be a close estimate of the total shear stress $\tau$ applied on the system:
\begin{equation}
\tau = \eta_f \frac{U - \langle V \rangle |_{z = z_s}}{h_f} \;
\end{equation}
with  $\langle V \rangle |_{z = z_s}$ the mean particle velocity measured at $z_s$. Our calculation of $\tau$ differs from previous studies\citep{Charru2004, Ouriemi2009, RevilBaudard2015} in that we define the bed surface from the concentration profile, and that we take into account the slip velocity of particles at the surface. For sediment transport studies it is common to normalize the shear stress by a normal stress due to the particle weight, to compute the Shields number:
\begin{equation}
\tau^* = \frac{\tau}{(\rho_p - \rho) g d} \;.
\end{equation} 
As we increase $\tau^*$ and the sediment transport rate increases, the width of the transition from the quasi-static bed to the fluid --- where the particle concentration drops --- becomes broader (see figure \ref{fig:CIvProfiles}a). 



\begin{figure}
\centerline{\includegraphics[width=300pt]{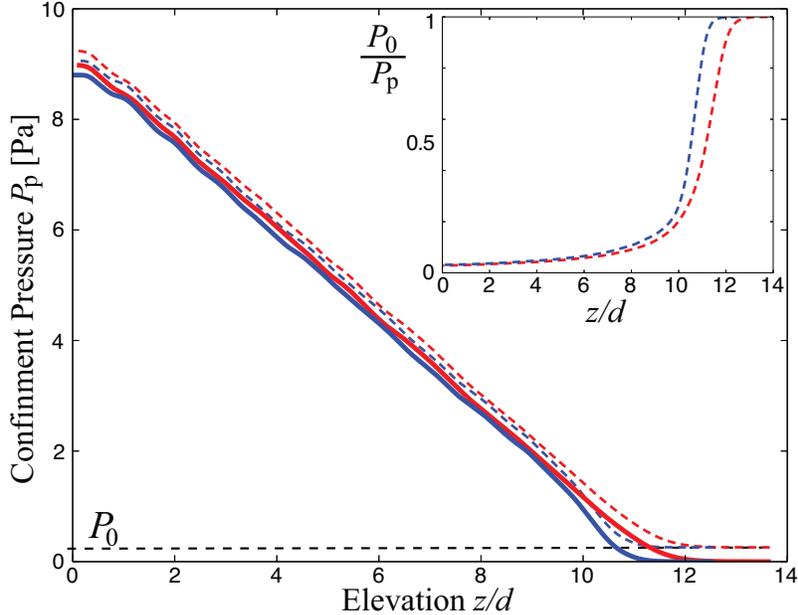}}
\caption{vertical profile of the confinement pressure $P_p$ in the system for $U = 14~\mathrm{mm/s}$ (blue) and $U = 48~\mathrm{mm/s}$ (red). Plain lines are the profiles obtain for $P_0 = 0$, and dashed lines are the profiles for $P_0 = 0.25~\mathrm{Pa}$. Insert: vertical profile of the ratio of $\frac{P_0}{P_p}$, using $P_0 = 0.25~\mathrm{Pa}$. }
\label{fig:PpProfiles}
\end{figure}

\subsection{Rheology using $P_0 = 0$}
As discussed in the Introduction, a major difference between the local rheology model developed and applied in the experimental system of Boyer \textit{et al.}\citep{Boyer2011}, and a sediment transport system with a free surface, is the treatment of the pressure. Instead of a constant confining pressure applied from the container, there is a depth-varying pressure that results from the weight of the grains. Previous experimental sediment-transport studies have employed the local rheology model with depth-varying pressure\citep{Aussillous2013,Revil2015}, but they did not include the pressure term $P_0$ proposed here. To understand the significance of this additional term, we examine granular rheology first by assuming $P_0 = 0$, the simplest hypothesis. In the next subsection we compare these findings to results that include a non-zero $P_0$ value.
Figure \ref{fig:CIvProfiles}a shows the profiles of Viscous number $I_v$, computed from the packing fraction and velocity profiles. Interestingly, on one hand, as already showed by \citet{Houssais2015}, the velocity kink deep in the bed corresponds to a Viscous number kink at $I_v \simeq 10^{-7}$. But on an other hand, all the profiles appear to converge to $I_v \simeq 1$ close to the surface. This observation is consistent with the expectation that a dynamical transition from the granular regime to the suspension regime occurs as $I_v$ approaches 1\citep{Boyer2011}. This is supported by the behavior of the effective viscosity $\eta_{eff}$, which saturates at high packing fraction at a value ($\simeq 10^{7} \eta_f$); in the other limit, all profiles converge to $\eta_{eff} \simeq 3 \eta_f$ at the surface as the concentration decreases toward zero (Fig. \ref{fig:CIvProfiles}b). Notably, for the the two highest stresses $\eta_{eff}$ profiles continue to decay toward $\eta_{f}$; i.e, the effective viscosity is determined only by the fluid. 
Taken together, data show the appropriate limits and indicate that sediment transport undergoes a transition from a dense granular material to a suspension.

\begin{figure}
\centerline{\includegraphics[width=450pt]{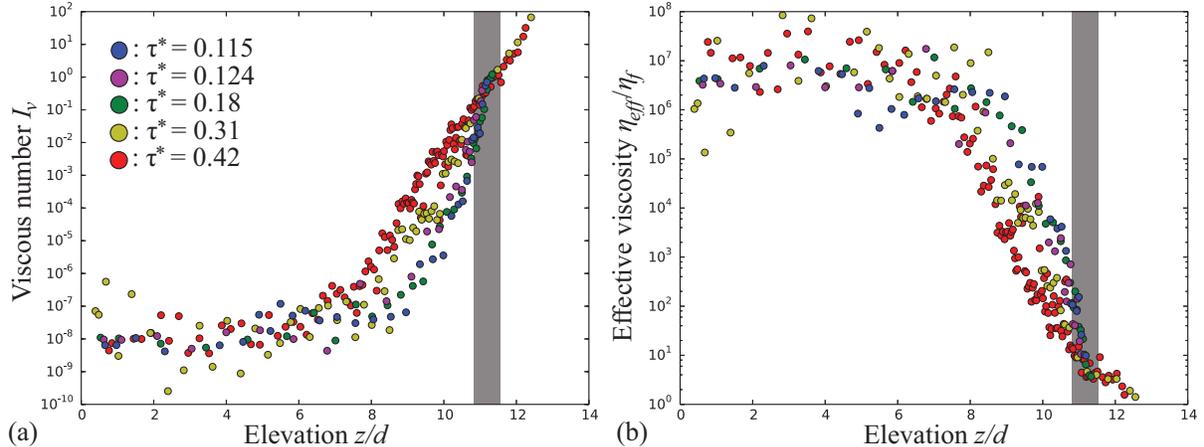}}
\caption{Vertical profiles of long-time averaged a) viscous number and b) effective viscosity. The gray area represents the range of elevations where concentration profiles $\langle \Phi \rangle (z) = \langle \Phi_{sat} \rangle_m/2$, an estimate of the bed surface locations. }
\label{fig:CIvProfiles}
\end{figure}

\begin{figure}
\centerline{\includegraphics[width=450pt]{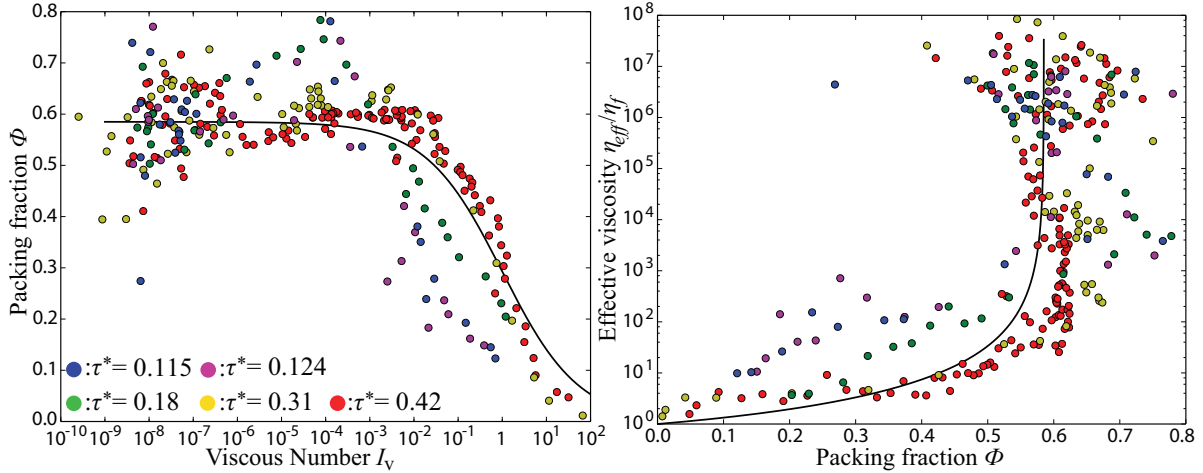}}
\caption{ a) Long-time averaged particle concentration as a function of the Viscous Number (computed using the confining pressure $P_p$). Black line represents equation (\ref{eq:CIv}). b) Long-time averaged effective viscosity $\eta_{eff}$ as a function of the concentration. Black line represents the effective viscosity relationship with packing fraction resulting from equations (\ref{eq:muIv}) and (\ref{eq:CIv}).}
\label{fig:CVsIv}
\end{figure}

\begin{figure}
\centerline{\includegraphics[width=300pt]{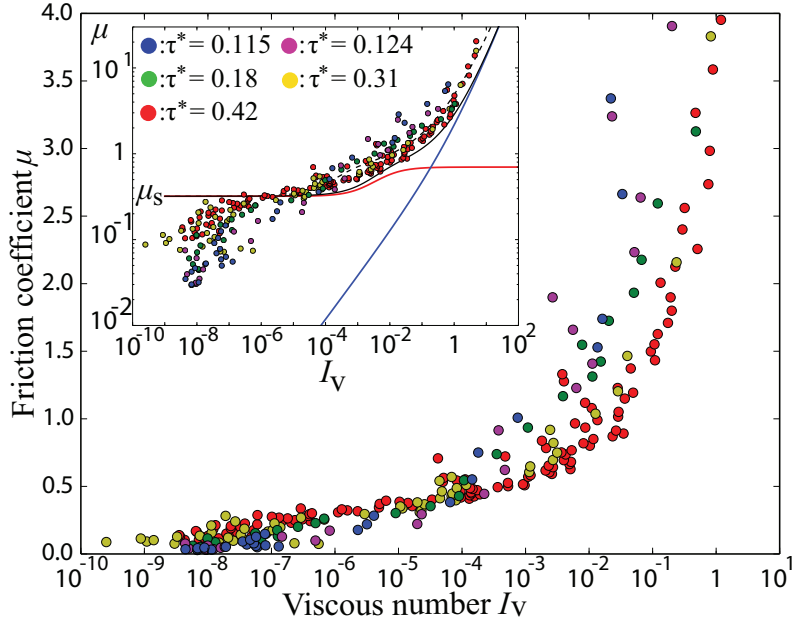}}
\caption{Friction coefficient $\mu$ as function of $I_v$, computed using $P_0 = 0$. Insert presents the logarithm of $\mu$, where the red and blue lines represent $\mu_{dry}(I_v)$ and $\mu_{susp}(I_v)$ respectively, and the black line represents equation (\ref{eq:muIv}). }
\label{fig:MuVsIv}
\end{figure}

On figures \ref{fig:CVsIv}  and \ref{fig:MuVsIv} we confront our dimensionless data of $\Phi$, $I_v$, $\eta_{eff}$ and $\mu$ with the extended local granular rheology proposed by Boyer \textit{et al.}\citep{Boyer2011}. On figure \ref{fig:CVsIv} the data are broadly consistent with the model, as they cluster around the relations from equations (\ref{eq:muIv}) and (\ref{eq:CIv}). 
Nonetheless, the data do not exhibit a clear collapse for the packing fraction relations $\langle \Phi \rangle(I_v)$ or $\eta_{eff}(\langle \Phi \rangle)$, except deep inside the bed. Instead, it appears that the lower the Shields number $\tau^*$ is, the more our results deviate from the model. The deviation is even more apparent for the local effective friction coefficient $\mu = \tau/P_p(I_v)$ (Fig. \ref{fig:MuVsIv}). For values $I_v > 10^{-5}$, the deviation of $\mu$ from the rheology model becomes more severe as the Shields number decreases. Moreover, $\mu$ grows to unphysically large values. We interpret this effect to be the result of a systematic bias, that arises due to the lack of the constant pressure $P_0$. 



\subsection{Rheology using $P_0 = \tau_c/\mu_s$}

Using $P_0=0$ makes the continuity of the sediment transport rheology problematic, and is not supported by our local rheology measurements. Because particles are settling, there exists a finite elevation where the packing fraction drops to zero (Fig. \ref{fig:SedTransportSketch}a). Accordingly, if $P_0 = 0$, the viscous number diverges to infinity in the dilute region ($\langle \Phi \rangle <<1$). However, dilute (or rarefied\citep{Furbish2012b}) particle transport near the onset of motion doesn't correspond to a gas, as the granular rheology predicts; rather, behavior is closer to that of a single particle on a quasi-static bed. We propose that the term $P_0$ be constructed such that the local flow rheology model converges to this asymptotic behavior (figure \ref{fig:SedTransportSketch}b); i.e., $\mu$ approaches $\mu_s$ at the bed surface as $\tau$ approaches $\tau_c$\citep{Shields1936, Wiberg1987}. 
Therefore, we propose:
\begin{equation}
P_0 = \frac{\tau_c}{\mu_s} \; .
\label{eq:defP0}
\end{equation}
As our experiments have been performed with PMMA spherical particles, similar to those used by Boyer \textit{et al.}\citep{Boyer2011}, we used the same value for $\mu_s = 0.32$ to compute $P_0$. Determining $\tau_c$ involves some ambiguity, however, as its value is known to vary widely with the measurement method \citep{Buffington1997}. We tested different $\tau_c$ values, and found a reasonably good collapse of the data for $ 0.06 ~\mathrm{Pa} \leq \tau_c \leq 0.1~\mathrm{Pa}$, which corresponds to a range of critical Shields number $0.03 \leq \tau^*_c \leq 0.05$. The data presented in subsequent figures are computed with $P_0 = 0.258~\mathrm{Pa}$, using $\tau^*_c = 0.04$. This value of $P_0$ is physically meaningful as it is of the same order of magnitude as the normal stress of a single particle, $P_0 \simeq 0.2 \times (\rho_p - \rho) g d/3$. The effect of $P_0$ on the computed pressure profile is negligible deep in the bed, but becomes more significant on approach to the surface (Fig. \ref{fig:PpProfiles}). Whereas the pressure $P_p$ converges to zero at vanishing $\langle \Phi \rangle$ without $P_0$, including this term results in the pressure saturating to $P_p = P_0$; this has a considerable impact on the computed pressure in the vicinity of the bed surface (see figure \ref{fig:PpProfiles} insert).

\begin{figure}
\centerline{\includegraphics[width=300pt]{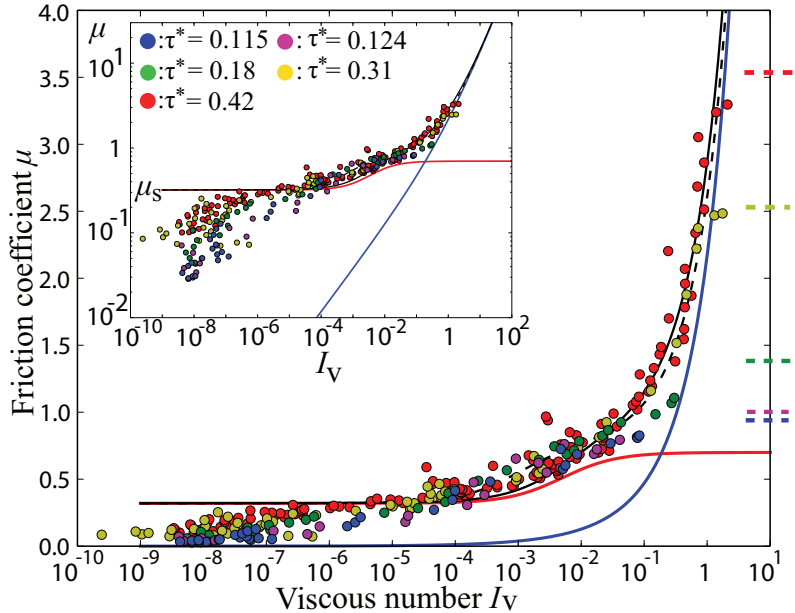}}
\caption{Friction coefficient $\mu$ as function of $I_v$, computed using the confining pressure $P_0 = \tau_c/\mu_s$. Insert presents the logarithm of $\mu$. The red and blue lines represent $\mu_{dry}(I_v)$ and $\mu_{susp}(I_v)$ respectively, and the black line represents equation (\ref{eq:muIv}). The black dash line represents a best fit result using equation (\ref{eq:muIv}), where $\mu_s$ and $\mu_d$ are the only fixed parameters (fit gives: $I_0 = 0.001$ and  $\Phi_c = 0.45$). The color dashed lines on the right represent the asymptotic values of $\mu$ in clear fluid, when $P_p = P_0$.}
\label{fig:MuVsIvWithP0}
\end{figure}

Including the pressure term $P_0 = 0.258~\mathrm{Pa}$ has several important consequences. First, data from all experiments collapse onto a single $\mu (I_v)$ curve for $I_v \geq 10^{-5}$ (Fig. \ref{fig:MuVsIvWithP0}). In other words, the rheology becomes independent of Shields stress. Second, these data cluster very close to the local rheology model prediction over the range of collapse. Third, the friction coefficient does not diverge indefinitely in the high $I_v$ limit. Instead, the profiles for each experiment depart from the theoretical curve as $P_p$ approaches $P_0$, and converge toward a finite value of the dynamic friction coefficient $\mu = \tau/P_0$ associated with the clear-fluid limit $\Phi = 0$.

A remarkable finding is that, for $I_v < 10^{-5}$, the data do not show convergence of the friction coefficient with the static value ($\mu = \mu_s$). Instead, $\mu$ decays continuously below $\mu_s$ with decreasing $I_v$, and the different experimental curves deviate from each other for values $\mu < \mu_s$. We do not observe any saturation of these trends.  

\section{Discussion}

In the Boyer \textit{et al.}\citep{Boyer2011} experiment, a single flow state associated with a single bulk viscous number was observed at a time, under an imposed confining pressure and packing fraction, and for the range: $10^{-6}<I_v<0.2$. For our system, the local shear rate and packing fraction adjust dynamically to the imposed fluid stress because of the free-surface condition. This results in a depth-varying viscous number and, as a result, multiple flow regimes coexist over the range: $10^{-9}<I_v<1$. Despite these differences, we find that the $\mu(I_v)$ rheology proposed by Boyer \textit{et al.}\citep{Boyer2011} can be extended to the case of settling particles sheared from above by a fluid, with the addition of a physically motivated pressure term $P_0 = \tau_c/\mu_s$ that accounts for particle weight. Our data closely follow the modified model for the range $10^{-5} \leq I_v \leq 1$. This result shows that a single rheology is capable of describing the complex case of sediment transport from bed load to suspension, as a transition from a slow and dense to a fast and dilute granular flow. This sediment transport regime is bounded by a fluid flow above where particle concentration vanishes, and a creeping granular system below where the local rheology model breaks down. 

The success of the pressure term $P_0$ in collapsing the data and recovering the rheology prediction confirms that this is a physically-meaningful term. Also, the inferred value for $t_{micro} = 0.28$s from $P_0$ is of the same order as the particle free-fall timescale $d/V_s = 1.2$s, where $V_s = g(\rho_p-\rho)d^2/(18\eta_f)$ is the Stokes velocity. This suggests a new method for assessing the critical stress $\tau_{c}$ from dynamics, which is quite different from the usual approach that extrapolates the flux-stress relation to zero. Our inferred value $\tau^*_c = 0.04$ is low relative to previous studies in laminar flow, where reported values are typically twice as large \citep{Charru2004, Ouriemi2007}. Nonetheless, it is compatible with the very sparse particle motion we observed at the surface during an experiment performed at $\tau^* = 0.043$ (see movie 3 in Supplementary information of \citet{Houssais2015}). It is also close the value reported by \citet{Charru2004} at the start of their experiments --- before any compaction occurred --- where entrainment of loose surface particles may approximate the situation of a single grain resting on the bed \ref{fig:SedTransportSketch}b.

It appears that the critical condition for motion of an individual particle on the bed surface may be characterized by a static friction threshold through $\tau^*_c$. The $\mu(I_v)$ rheology, however, indicates that $\tau^*_c$ does not represent a well-defined yield stress criterion for the granular system. Our results support recent studies calling for a modification to the classical bed-load transport framework; in particular, that the friction coefficient cannot be considered constant\citep{Ouriemi2009,Duran2012}, and that the ``bed-load active layer" is not constant but instead expands vertically in both directions as the shear stress $\tau$ increases \citep{Duran2012}. Results also inform models for suspended-sediment transport, supporting the idea of Boyer \textit{et al.}\citep{Boyer2011} that particle-particle frictional interactions should be taken into account, even for smooth spheres suspended in a viscous fluid. Interestingly, a similar reasoning has been developed recently in order to understand shear thickening in suspensions\citep{Wyart2014, Mari2014}. This suggests that improvements in our understanding of particle-scale interactions may yield a local rheology model capable of describing granular flow across all configurations of packing fraction and shear stress.

Recent studies have proposed to model dense sediment transport using $\Phi(I_v)$, and in particular have utilized a closure scheme in which granular transport ceases at a critical packing fraction \citep{Trulsson2012, Chiodi2014}. Our results, however, indicate that the relationship $\Phi(I_v)$ (equation (4)) is irrelevant for describing sediment transport in the vicinity of the onset of motion. This raises the issue then of how to predict the packing fraction profile, which was only measured for our experiments. Suspension modeling studies have proposed that particle diffusion due to internal pressure induces a flux normal to the shear \citep{Leighton1987, Boyer2011}, and that this diffusion mechanism may be used to capture sediment transport dynamics\citep{Chiodi2014}. This remains to be validated, however. To be relevant for sediment transport, we propose that closure equation for the concentration profile should also be consistent with the condition of a quasi-static bed at $\tau^* = \tau^*_c$.

Finally, our results identify two regime transitions where the local rheology breaks down. The first is the very dilute regime where $P_p$ approaches $P_0$, but this limit concerns very few particles which --- due to sedimentation --- tend to return to the denser part of the flow. Indeed, the deviation from the rheology relation in the large $I_v$ limit is modest (see figure \ref{fig:MuVsIvWithP0}); and, this deviation represents a physically meaningful limit that the friction approaches the dynamic friction value for a single particle. The second, more important transition occurs for $I_v< 10^{-5}$, where data from different experiments deviate from the model and each other. Based on observations of particle motion\cite{Houssais2015}, we interpret this deviation as the signature of the creep regime. Creep occurs for values $\mu<\mu_s$, where $\mu$ continues to decline with decreasing $I_v$. The local rheology model predicts that $\mu$ converges to $\mu_s$ for vanishing $I_v$, representing a jamming transition \cite{GDRMidi2004,Boyer2011, Chiodi2014}. Our experiments, which reach values for $I_v$ more than three orders of magnitude smaller than reported in Boyer \textit{et al.}\citep{Boyer2011} study, do not show any jamming transition. It is important to note that creep is associated with localized, intermittent particle motion such that the average particle velocity profile -- and so $I_v$ -- may be irrelevant for describing particle dynamics in the creeping regime. It is possible that the assumption of strictly local particle interactions is broken for creep, where collective particle motion may be expected to occur\citep{Katsuragi2010}. Future examination of dynamical heterogeneities in the experimental data will help to address this issue. From the theoretical side, the recent development of a non-local rheology framework\citep{Lerner2012, Bouzid2013,Kamrin2015} is a promising approach for modeling creep dynamics. Currently, model predictions appear to be inconsistent with our observations\citep{Kamrin2012}, as they predict deviation from the local rheology at large Viscous number. Results cannot be directly compared at present, however, as the non-local models implement a boundary condition of no particle motion very far from the shear zone. For the few experiments where creep has been quantified, boundary slip has occurred\citep{Nichol2010,Reddy2011,Houssais2015}. Modification of non-local rheology models to incorporate the geometry and boundary conditions of experiments will allow for a direct test.


%
%

\section{Conclusion}

Using a novel experimental setup and protocol, we capture the rheology of sediment transport near the onset of particle motion. By accounting for the asymptotic quasi-static behavior of particles at the surface for $\tau = \tau_c$, we link the classical definition of critical shear stress $\tau_c$ to the local rheology of a granular flow submerged in a viscous fluid. 
These results provide a new perspective on the modeling of sediment transport processes with continuum mechanics, and open the possibility that creeping to suspension regimes --- which are responsible for most of landscape dynamics --- may be described with a unified rheology.   
Our results emphasize the importance of the pressure $P_0$ at the bed surface, corresponding to the weight of an individual particle. This parameter may be relevant for transport and segregation of mixed grain sizes in submerged granular flows.
Finally, the transition to creep at low viscous number challenges our understanding of local rheology and the nature of the jamming transition. The motion of particles in the creep regime deserves more attention, which may motivate new comparisons with non-local rheology models. Many rivers and hillslopes are granular systems that self-organize such that they are in the vicinity of the threshold of motion \cite{Jerolmack2011}. Thus, a better understanding of creep dynamics will improve long-term predictions of landscape evolution.

\bibliography{RheologyBib}

\end{document}